\documentclass[showpacs,preprint]{revtex4}

\usepackage{graphicx}
\usepackage{dcolumn}
\usepackage{amsmath}

\begin{document}


\title{Exact EGB models for spherical static perfect fluids}


\author{Sudan Hansraj} \email[]{hansrajs@ukzn.ac.za}
\author{Brian Chilambwe} \email[]{brian@aims.ac.za}
\author{Sunil D. Maharaj} \email[]{maharaj@ukzn.ac.za}
\affiliation{Astrophysics and Cosmology Research Unit, School of Mathematics, Statistics and Computer Science, University of KwaZulu--Natal, Private Bag X54001, Durban 4000, South Africa}


\date{\today}

\begin{abstract}

We obtain a new exact solution to the field equations in the EGB modified theory of gravity for a 5--dimensional spherically symmetric static distribution. By using a transformation, the study is reduced to the analysis of a single second order nonlinear differential equation. In general the condition of pressure isotropy produces a first order differential equation which is an Abel equation of the second kind. An exact solution is found. The solution is examined for physical admissability.  In particular a set of constants is found which ensures that a pressure--free hypersurface exists which defines the boundary of the distribution. Additionally the isotropic pressure and the energy density are shown to be positive within the radius of the sphere. The adiabatic sound speed criterion is also satisfied within the fluid ensuring a subluminal sound speed. Furthermore, the weak, strong and dominant conditions hold throughout the distribution. On setting the Gauss--Bonnet coupling to zero, an exact solution for 5--dimensional perfect fluids in the standard Einstein theory is obtained. Plots of the dynamical quantities for the Gauss--Bonnet and the Einstein case reveal that the pressure is unaffected while the the energy density increases under the influence of the Gauss--Bonnet term.

\end{abstract}


\maketitle


\section{Introduction}

Alternate or extended theories of gravity have aroused considerable interest recently in view of difficulties with the general theory of relativity to explain anomalous behaviour of gravitational phenomena such as the late time expansion of the universe. The mathematical reason for this interest is that the higher order derivative curvature terms make a nonzero contribution to the dynamics. In particular, Einstein--Gauss--Bonnet (EGB) theory has proved promising in this regard, and therefore is the most extensively studied. It appears in a natural way in the effective action of heterotic string theory in the low energy limit \cite{1}. Several new results have been reported especially dealing with the aspect of gravitational collapse. The causal structure of the singularities is different from general relativity for inhomogeneous distribution of dust and null dust \cite{2}.

Historically black hole models in EGB theory have been intensively studied. Boulware and Deser \cite{3} generalized the higher dimensional solutions in Einstein theory due to Tangherlini \cite{4}, and by Myers and Perry \cite{5} to include the contribution of the EGB theory with quadratic curvature terms. Wheeler \cite{6}, Torii and Maeda \cite{7} and Myers and Simons \cite{8} have also considered black hole solutions in EGB theory. Inhomogeneous collapse of dust, that is pressure--free fluid with non--interacting particles, was studied by Maeda \cite{9}. However explicit exact solutions were obtained by Jhingan and Ghosh \cite{10}. Dadhich et al \cite{11} proved that the constant density Schwarzschild interior solution is universal in the sense that it is valid in both higher dimensional Einstein theory as well as in EGB gravity. The matching of these exterior metrics to an interior for brane world stars to produce analytical models was investigated by Casadio and Ovalle \cite{12}. The matching of isolated masses to a Schwarzschild exterior was analysed by Clifton et al \cite{13}.

To date, there appears to exist no {\it complete} stellar model in 5-dimensional EGB theory for the perfect fluid configuration of static spherically symmetric matter. The constant density configuration of Dadhich et al \cite{11} and the static spherically symmetric star of Kang et al \cite{14} need to satisfy the junction conditions of EGB gravity so that matching is possible at the stellar surface. The junction conditions for EGB were derived in Davis \cite{15} which are nontrivial and very different from general relativity. Note that the variable density model of Kang et al \cite{14} requires a further integration to produce an exact solution. As far as we are aware there is no known interior variable density spherically symmetric exact solution to the EGB field equations. This is the object of our study in this paper and to highlight the role of junction conditions of \cite{15}. The problem of finding an exact interior metric in EGB theory for spherically symmetric distribution reduces to solving a system of three nonlinear partial differential equations in four unkowns: the dynamical quantities pressure and energy density and two gravitational potentials. As the system is under--determined, it is necessary to specify an additional contraint in order to close the system. This process is analogous to the standard Einstein gravity case. Traditionally, the approach has been to specify one of the four unknowns and by integration to resolve the remaining three. The reader is referred to the comprehensive listing of exact solutions found in this way by Stephani et al \cite{16} and Delgaty and Lake \cite{17} for the Einstein case. When one of the field equations is replaced by the equation of hydrodynamic equilibrium, namely the vanishing divergence of the energy momentum tensor, then it is prudent to invoke an equation of state relating the energy density and pressure. This appears to have been the approach of \cite{14} in their attempt to find an exact model. Interestingly the interior model presented in \cite{14} generates the well known vacuum metric \cite{3} of EGB in the limit of vanishing pressure and density.

Recently Izaurieta and Rodriguez \cite{a0} argued that four dimensional gravity may effectively emerge from 5-dimensional EGB theory. 
The addition of diffeomorphism-invariant terms to the action principle leads to second order equations of motion and are therefore physically palatable.
Consequently investigations  in five dimensions are laboratories for examining the impact of extra dimensions on physics, for example by demonstrating that a new
 exact solution satisfies elementary physical properties demanded of astrophysical objects. 
In  four dimensions the higher order curvature terms in EGB theory do not affect gravity. It is only with spacetime dimensions five or greater when the 
Gauss-Bonnet term contributes nontrivially to the dynamics. The simplest case in higher dimensions is five which has been extensively studied in several
physical scenarios. We point out that the addition of the extra spatial dimension has been investigated  by Kang et al \cite{14} in static stars,  Brihaye and Reidel \cite{a1} in rotating boson stars,
Ghosh et al  \cite {2} in spherical collapsing bodies, and Chervon et al \cite{a2} in emergent universe models. The presence of additional dimensions may have
a dramatic effect on the behavior of matter. For example Maeda \cite{9} showed that massive timelike naked singularities may exist in five dimensions and massless ingoing
null naked singularities are formed in dimensions greater than five in EGB theory. Massive timelike naked singularities do not exist in general relativity.
Also note that the dynamics of  charged radiating gravitational collapse of shear-free matter has been studied recently in the socalled modified Gauss-Bonnet gravity \cite{a3}.

Our intention in this paper is to solve the nonlinear EGB equations for a static spherically symmetric matter distribution. In section II we briefly outline the basic equations in EGB gravity. The field equations in 5--dimensional EGB gravity are presented for a spherically symmetric metric, and they are then transformed to an equivalent form in Section III. An exact solution to the EGB equations is found in Section IV. In Section V a corresponding exact solution, in the Einstein case, to the 5--dimensional case is presented. The physical features of the model are investigated in Section VI. Some concluding remarks are made in Section VII. In the Appendix we present several exact solutions in the Einstein case in five dimensions by specifying a form for one of the metric potentials.

\section{Einstein--Gauss--Bonnet Gravity}

The Gauss--Bonnet action in five dimensions is written as
\begin{equation}
S = \int \sqrt{-g} \left[ \frac{1}{2} \left(R - 2\Lambda + \alpha L_{GB}\right)\right] d^5 x + S_{\mbox{ matter}}, \label{1}
\end{equation}
where $ \alpha $ is the Gauss--Bonnet coupling constant. The strength of the action $ L_{G B} $ lies in the fact that despite the Lagrangian being quadratic in the Ricci tensor, Ricci scalar  and the Riemann tensor, the equations of motion turn out to be second order quasilinear which are compatible with a theory of gravity. The Gauss--Bonnet term is of no consequence for $ n \leq 4 $ but is generally nonzero for $ n > 4 $.

The EGB field equations may be written as
\begin{equation}
G_{a b} + \alpha H_{a b} = T_{a b},  \label{2}
\end{equation}
with metric signature $ (- + + + +) $ where $ G_{ab} $ is the Einstein tensor. The Lanczos tensor is given by
\begin{equation}
H_{a b} = 2 \left(R R_{a b} - 2 R_{a c}R^{c}_{b} - 2 R^{c d} R_{a c b d} + R^{c d e}_{a} R_{b c d e} \right) - \frac{1}{2} g_{a b} L_{G B},  \label{3}
\end{equation}
where the Lovelock term has the form
\begin{equation}
L_{G B} = R^2 + R_{a b c d} R^{a b c d} - 4R_{c d} R^{c d}.   \label{4}
\end{equation}

\section{Field Equations}

The generic 5--dimensional  line element for static spherically symmetric spacetimes is taken as
\begin{equation}
ds^{2} = -e^{2 \nu} dt^{2} + e^{2 \lambda} dr^{2} + r^{2} \left( d\theta^{2} + \sin^{2} \theta d \phi + \sin^{2} \theta \sin^{2} \phi d\psi \right), \label{5}
\end{equation}
where $ \nu(r) $ and $ \lambda(r) $ are  the gravitational potentials. We utilise a comoving fluid velocity of the form $ u^a = e^{-\nu} \delta_{0}^{a} $ and the matter field is that of a perfect fluid with energy momentum tensor $ T_{a b} = (\rho + p) u_a u_b + p g_{a b} $. Accordingly the EGB field equations (\ref{2})  reduce to
\begin{eqnarray}
\rho &=& \frac{3}{e^{4 \lambda} r^{3}} \left( 4 \alpha \lambda ^{\prime} +  r e^{2 \lambda} -  r e^{4 \lambda} -  r^{2} e^{2 \lambda} \lambda ^{\prime} - 4 \alpha e^{2 \lambda} \lambda ^{\prime} \right),  \label{6a} \\ \nonumber \\
p &=&  \frac{3}{e^{4 \lambda} r^{3}} \left(-  r e^{4 \lambda} + \left( r^{2} \nu^{\prime} +  r + 4 \alpha \nu^{\prime} \right) e^{2 \lambda} - 3 \alpha \nu^{\prime} \right),  \label{6b} \\ \nonumber \\
p &=& \frac{1}{e^{4 \lambda} r^{2}} \left( -e^{4 \lambda} - 4 \alpha \nu^{\prime \prime} + 12 \alpha \nu^{\prime} \lambda^{\prime} - 4 \alpha \left( \nu^{\prime} \right)^{2}  \right) \nonumber \\
                 & \quad & + \frac{1}{e^{2 \lambda} r^{2}} \left(  1 - r^{2} \nu^{\prime} \lambda^{\prime} + 2 r \nu^{\prime} - 2 r \lambda^{\prime} + r^{2} \left( \nu^{\prime} \right)^{2}  \right) \nonumber \\
                 & \quad & + \frac{1}{e^{2 \lambda} r^{2}} \left(  r^{2} \nu^{\prime \prime} - 4 \alpha \nu^{\prime} \lambda^{\prime} + 4 \alpha \left( \nu^{\prime} \right) ^{2} + 4 \alpha \nu^{\prime \prime}   \right). \label{6c}
\end{eqnarray}
Note that the system (\ref{6a})--(\ref{6c}) comprises three field equations in four unknowns which is similar to the standard Einstein case for spherically symmetric perfect fluids. Observe that the  vacuum metric describing the gravitational field exterior to the 5--dimensional static perfect fluid may be described by the Boulware--Deser \cite{3} spacetime as
\begin{equation}
ds^2 = - F(r) dt^2 + \frac{dr^2}{F(r)} + r^{2} \left( d\theta^{2} + \sin^{2} \theta d \phi + \sin^{2} \theta \sin^{2} \phi d\psi \right), \label{7}
\end{equation}
where
\[
F(r) = 1 + \frac{r^2}{4\alpha} \left( 1 - \sqrt{1 + \frac{8M\alpha}{r^4}} \right).
\]
In the above $ M $ is associated with the gravitational mass of the hypersphere. The exterior solution is not unique and neither is there a Birkhoff theorem analogous to the 4--dimensional gravity case. At least this metric involves branch cuts. Bogdanos et al \cite{18} have analysed the 6--dimensional case in EGB and demonstrated the validity of Birkhoff's theorem for this order.

We invoke the transformation $ e^{2 \nu} = y^{2}(x) $, $ e^{-2 \lambda} = Z(x)  $ and $ x = C r^{2} $ ($ C $ being an arbitrary constant) which was utilised successfully by Durgapal and Banerji \cite{19}, Finch and Skea \cite{20} and Hansraj and Maharaj \cite{21} to generate new exact solutions for neutral and charged isotropic spheres. For applications to charged anisotropic relativistic matter with this transformation see the recent works of Mafa Takisa and Maharaj \cite{22} and Maharaj et al \cite{23} in four dimensions. The field equations (\ref{6a})--(\ref{6c}) may be recast as
\begin{eqnarray}
3  \dot{Z} + \frac{3  (Z - 1) ( 1 - 4 \alpha C \dot{Z} )}{x} &=& \frac{\rho}{C}, \label{8a} \\ \nonumber \\
\frac{3  (Z - 1)}{x} + \frac{6  Z \dot{y}}{y} - \frac{24 \alpha C (Z - 1) Z \dot{y}}{x y} &=& \frac{p}{C}, \label{8b} \\ \nonumber \\
 2 x Z \left( 4 \alpha C [Z - 1] - x \right) \ddot{y}  - \left( x^{2} \dot{Z} + 4 \alpha C \left[ x \dot{Z} - 2 Z + 2 Z^{2} - 3 x Z \dot{Z} \right] \right) \dot{y} \nonumber \\ - \left( 1 + x \dot{Z} - Z \right) y &=& 0,  \label{8c}
\end{eqnarray}
where the last equation is called the equation of pressure isotropy. Equation (\ref{8c}) has been arranged as a second order differential equation in $ y $, which for some analyses in the 4--dimensional Einstein models, proves to be a useful form. Functional forms for $ Z(x) $ may be selected {\it a priori} so as to allow for the complete integration of the field equations. For example, the form $ Z=1+x $ produces a higher dimensional Schwarzschild solution with constant density. This corroborates the result of Dadhich et al \cite{11} that the constant density Schwarzschild solution is universal - that is it is independent of dimension. We have also found a number of other cases for $ Z $ for which (\ref{8c}) is integrable and these will be dealt with in the future.

For the present work it should be noted that (\ref{8c}) may also be regarded as a first order ordinary differential equation in $ Z $, and may be expressed in the form
\begin{eqnarray}
\left( x^{2} \dot{y} + x y + 4 \alpha C x \dot{y} - 12 \alpha C x \dot{y} Z \right) \dot{Z} + 8 \alpha C \left( \dot{y} - x \ddot{y} \right) Z^{2} \nonumber \\+ \left( 2 x^{2} \ddot{y} + 8 \alpha C x \ddot{y} - 8 \alpha C \dot{y} - y \right) Z + y = 0. \label{9}
\end{eqnarray}
This is an Abel equation of the second kind for which few solutions are known. However note that by choosing forms for $ y $ should in theory result in the expressions for $ Z $ by integration. Therefore we seek choices for the metric potential $ y $ which will allow for a complete resolution of the geometrical and dynamical variables.

\section{New Exact Interior Solution in the EGB case}

Locating exact solutions for (\ref{9}) is difficult to achieve in view of its nonlinearity.  One strategy is to investigate the consequence of one or more of the coefficients to vanish. In requiring that the coefficient of $ Z^2 $ vanishes we obtain the restriction
\begin{equation}
\dot{y} - x \ddot{y} = 0, \label{10}
\end{equation}
which may be solved to give
\begin{equation}
y = \frac{1}{2} C_{1} x^{2} + C_{2}, \label{11}
\end{equation}
where $ C_{1} $ and $ C_{2} $ are constants of integration. Note that the restriction (\ref{10}) simplifies (\ref{9}) but does not remove its nonlinearity. Inserting (\ref{11}) into (\ref{9}) the condition of pressure isotropy is transformed to
\begin{equation}
\left[ 3 C_{1} x^{3} + 8 \beta C_{1} x^{2} + 2 C_{2} x - 24 \beta C_{1} x^{2} Z \right] \dot{Z} + \left[ 3 C_{1} x^{2} - 2 C_{2} \right] Z + C_{1} x^{2} + 2 C_{2} = 0, \label{12}
\end{equation}
where for convenience we set $ \beta = \alpha C $. Renaming $ \frac{C_1}{C_2} = \epsilon $ equation  (\ref{12}) assumes the simpler form
\begin{equation}
\left[ 3 \epsilon x^{3} + 8 \beta \epsilon x^{2} + 2  x - 24 \beta \epsilon x^{2} Z \right] \dot{Z} + \left[ 3 \epsilon x^{2} - 2  \right] Z + \epsilon x^{2} + 2  = 0, \label{13}
\end{equation}
which will aid our graphical investigations. The parameters $ \epsilon $ and $ \beta $ will have to be assigned values to obtain the qualitative features of the eventual model. On solving  (\ref{13}), we obtain the solutions
\begin{equation}
Z = \frac{3 \epsilon x^{2} + 8 \beta \epsilon x + 2 \pm \Upsilon}{24 \beta \epsilon x}, \label{14}
\end{equation}
where \[ \Upsilon = \sqrt{3 \epsilon ^{2}(3 x^{4} + 32 \beta x^{3}) + 4 \epsilon \left( 16 \beta^{2} \epsilon + 3 + 144 \beta^{2} \epsilon C_{3} \right) x^{2} - 4 (16 \beta \epsilon x - 1)}, \] and $ C_{3} $ is an integration constant.

With the help of  $ (\ref{14}) $ and $ (\ref{8a}) $  the energy density for the EGB case is given by
\begin{eqnarray}
\frac{\rho}{C} &=& \frac{27 \epsilon ^{2} x^{4} - 48 \beta \epsilon ^{2} x^{3} - 32 \beta ^{2} \epsilon x + 4 - 4 \left( 1 - 4 \beta \epsilon x \right) \Upsilon  }{48 \beta \epsilon ^{2} x^{4}} \nonumber \\
               & \quad & - \frac{ x \left( 3 \epsilon x^{2} + 16 \beta \epsilon x - 2 \right) \Upsilon ^{\prime} - \left( \Upsilon - x \Upsilon ^{\prime} \right) \Upsilon }{48 \beta \epsilon ^{2} x^{4}}, \label{15}
\end{eqnarray}
while the pressure $ p $  has the form
\begin{eqnarray}
\frac{p}{C} &=& \frac{ 27 \epsilon ^{2} x^{4} + 96 \epsilon ^{2} \beta x^{3} + 8 \epsilon \left( 3 + 32 \epsilon \beta ^{2} \right) x^{2} - 64 \epsilon \beta x + 4 }{24 \epsilon \beta x ^{2} (\epsilon x^{2} + 2)} \nonumber \\
& \quad & - \frac{ \left( 3 \epsilon x^{2} + 16 \epsilon \beta x - 2 + 2 \Upsilon \right) \Upsilon}{24 \epsilon \beta x (\epsilon x^{2} + 2)}, \label{16}
\end{eqnarray}
via (\ref{8b}).
The adiabatic sound speed parameter is found to be
\begin{equation}
\frac{d p}{d \rho} = \frac{2 \epsilon x^{2} U(x)}{ \left( \epsilon x^{2} + 2 \right) ^{2} V(x) }, \label{17}
\end{equation}
where $ U(x) $ and $ V(x) $ are, respectively, given by
\begin{eqnarray}
U(x) & = & - 96 \epsilon ^{3} \beta x^{5} + 4 \epsilon ^{2} \left( 15 - 128 \epsilon \beta ^{2} \right) x^{4} + 384 \epsilon ^{2} \beta x^{3} - 16 \epsilon x^{2} \nonumber \\
     & \quad & + 128 \epsilon \beta x - 16 + 8 \left( \epsilon x^{2} + 1 \right) \Upsilon ^{2} - 4 x \left( \epsilon x ^{2} + 2 \right) \Upsilon \Upsilon ^{\prime} \nonumber \\
     & \quad & + 2 \left( 3 \epsilon ^{2} x^{4} + 24 \epsilon ^{2} \beta x^{3} - 4 \epsilon x ^{2} + 16 \epsilon \beta x - 4 \right) \Upsilon \nonumber \\
     & \quad & - x \left( 3 \epsilon ^{2} x^{4} + 16 \epsilon ^{2} \beta x ^{3} + 4 \epsilon x ^{2} + 32 \beta \epsilon x - 4 \right) \Upsilon ^{\prime}, \nonumber
\end{eqnarray}
and
\begin{eqnarray}
V(x) & = & 48 \epsilon ^{2} x ^{3} + 96 \beta \epsilon x - 16 + 16 \left( 1 - 3 \beta \epsilon x \right) \Upsilon - 4 \Upsilon ^{2} \nonumber \\
     & \quad & + x \left( 3 \epsilon x^{3} + 48 \epsilon \beta - 10  \right) \Upsilon ^{\prime} + x \left( 5 \Upsilon + x \Upsilon ^{\prime} \right) \Upsilon ^{\prime} \nonumber \\
     & \quad & - x ^{2} \left( 3 \epsilon x ^{2} + 16 \epsilon \beta x - 2 + \Upsilon \right) \Upsilon ^{ \prime \prime}. \nonumber
\end{eqnarray}
To study the energy conditions we need to obtain forms for $ \rho - p $, $ \rho + p $ and $ \rho + 3 p $. Explicit forms for these expressions can be immediately generated from (\ref{15}) and (\ref{16}). We will compare these expressions with the corresponding forms of the 5--dimensional Einstein case later.

Other restrictions in (\ref{9}) may lead to new models in addition to that considered in this section. For example on setting the coefficient of $Z$ in (\ref{9}) to zero, we obtain
\begin{equation}
x \left( x + 1 \right) \ddot{y} - \dot{y} - \frac{1}{2} y = 0, \label{18}
\end{equation}
where we have set $ 4 \alpha C = 1 $ for simplicity. This produces a differential equation which is of the hypergeometric type, and consequently is not readily expressible in terms of elementary functions. Therefore the prospects of establishing $ Z $ explicitly are remote for the resultant form for $ y $. This case may be treated with the other methods and this will be considered later.  Additionally, it should be noted that the vanishing of the coefficient of $ \dot{Z} $ is not mathematically feasible as it involves both $ y $ and $ Z $.

\section{New Exact Interior Solution in the Einstein Case}

The 5--dimensional Einstein version of the above is obtained by setting $ \alpha = 0 $ in (\ref{9}) to give the differential equation
\begin{equation}
2x^2 Z \ddot{y} + x^2 \dot{Z} \dot{y} + (1-Z + \dot{Z}x) y =0. \label{19}
\end{equation}
This differs from its 4-dimensional counterpart
\begin{equation}
4x^2 Z \ddot{y} + 2x^2 \dot{Z} \dot{y} + (1-Z + \dot{Z}x) y =0, \label{20}
\end{equation}
in the coefficients of its first two terms  \cite{19}. Invoking the metric ansatz (\ref{11}), equation (\ref{19}) reduces to
\begin{equation}
x(3\epsilon x^2 + 1)\dot{Z} + (3\epsilon x^2 - 1) Z + (\epsilon x^2 + 1) = 0, \label{21}
\end{equation}
where we have set $ \tilde{\epsilon} = \frac{C_{1}}{ 2 C_{2}}$. The general solution to (\ref{21}) is given by
\begin{equation}
Z(x)=\frac{1 + C_{3} x - \tilde{\epsilon} x^2}{1 + 3 \tilde{\epsilon} x^2}. \label{22}
\end{equation}
It is important to observe that this does not follow as a special case of solution (\ref{14}) since $ \beta $ and consequently $ \alpha $ appears in the denominator. Now the dynamical quantities, pressure, energy density, sound speed index, and energy conditions have the forms
\begin{eqnarray}
p &=&  \frac{ 3 C \left[ C_{3} + 5 C_{3} \tilde{\epsilon} x^{2} - 8 \tilde{\epsilon} ^{2} x^{3} \right] }{(\tilde{\epsilon} x^{2} + 1) (1 + 3 \tilde{\epsilon} x^{2}) }, \label{23} \\
\rho &=&  \frac{ 6 C \left[ C_{3} - 6 \tilde{\epsilon} x - 6 \tilde{\epsilon} ^{2} x^{3}  \right] }{ (1 + 3 \tilde{\epsilon} x^{2})^{2} }, \label{24} \\
\frac{dp}{d\rho} &=& \frac{x (1 + 3 \tilde{\epsilon} x^{2} ) \left[ 12 \tilde{\epsilon} ^{3} x - 15 C_{3} \tilde{\epsilon} ^{2} x ^{4} - 16 \tilde{\epsilon} ^{2} x ^{3} - 6 C_{3} \tilde{\epsilon} x ^{2} - 12 \tilde{\epsilon} x + C_{3} \right] }{ 6 ( \tilde{\epsilon} x^{2} + 1)^{2} \left[ 3 \tilde{\epsilon} ^{2} x ^{4} + 6 \tilde{\epsilon} x ^{2} - 2 C_{3} x  - 1 \right] }, \label{25}
\end{eqnarray}

\begin{eqnarray}
\rho - p &=& \frac{3 C \left[ C_{3} - 12 \tilde{\epsilon} x - 6 C_{3} \tilde{\epsilon} x^{2} - 16 \tilde{\epsilon} ^{2} x ^{3} - 15 C_{3} \tilde{\epsilon} ^{2} x ^{4} + 12 \tilde{\epsilon} ^{3} x ^{5} \right] }{(1 + \tilde{\epsilon} x ^{2}) (1 + 3 \tilde{\epsilon} x ^{2})^{2}}, \label{26} \\
\rho + p &=& \frac{3 C \left[ 3 C_{3} - 12 \tilde{\epsilon} x + 6 C_{3} \tilde{\epsilon} x^{2} - 32 \tilde{\epsilon} ^{2} x ^{3} + 15 C_{3} \tilde{\epsilon} ^{2} x ^{4} - 36 \tilde{\epsilon} ^{3} x ^{5} \right] }{(1 + \tilde{\epsilon} x ^{2}) (1 + 3 \tilde{\epsilon} x ^{2})^{2}}, \label{27} \\ \nonumber \\
\rho + 3p &=& \frac{3 C \left[ 5 C_{3} - 12 \tilde{\epsilon} x + 26 C_{3} \tilde{\epsilon} x^{2} - 48 \tilde{\epsilon} ^{2} x ^{3} + 45 C_{3} \tilde{\epsilon} ^{2} x ^{4} - 84 \tilde{\epsilon} ^{3} x ^{5} \right] }{(1 + \tilde{\epsilon} x ^{2}) (1 + 3 \tilde{\epsilon} x ^{2})^{2}}, \label{28}
\end{eqnarray}
respectively. In the plots to follow, we exhibit the above quantities in comparison with their EGB counterparts to investigate the role of the Gauss--Bonnet term in the solution.

\section{Physical features}

We extrapolate the familiar conditions imposed on stellar configurations in the usual Einstein theory to the 5--dimensional EGB case. It would be interesting to see if the models generated in this modified theory of gravity satisfy these standard conditions. We require that the energy density $ \rho $ and pressure $ p $, and the metric potentials  $ e^{2 \nu} $ and $ e^{2 \lambda} $, should be regular in the interior. The radial pressure should vanish at the boundary $ r = R. $ The gradients $ \rho ^{\prime} $ and $ p^{\prime} $ should be negative for barotropic matter. The speed of sound should remain subluminal throughout the interior of the star. At the boundary $ r = R $ the metric functions should match smoothly to the exterior Boulware--Deser \cite{3} solution
\[
e^{2\nu(R)} = 1 + \frac{R^2}{4\alpha} \left( 1 - \sqrt{1 + \frac{8M\alpha}{R^4}} \right) =         e^{- 2\lambda (R)}.
\]
For realistic matter we require compliance with the energy conditions: weak energy condition $ ( \rho - p > 0 ) $, strong energy condition $ ( \rho + p > 0 ) $, and dominant energy condition $ ( \rho + 3p > 0 ) $.

Clearly a complete analytic treatment of our solution is ostensibly not possible given its complexity. From an examination of the pressure (\ref{16}) it is evident that solving for $ x $ in terms of $ p $ is not attainable as this reduces to solving an eighth degree polynomial equation, and there is no general way of doing this. This in turn means that it is not possible to write the density as a function of pressure to obtain a barotropic equation of state; the form $ p=p(\rho) $ is not possible.

\begin{figure}[h!]
  \includegraphics[width=8cm]{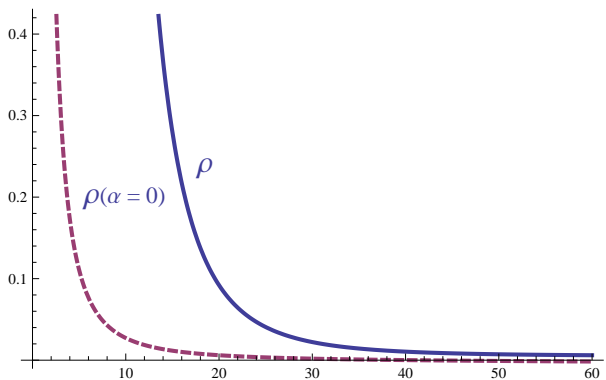}\\
  \caption{Plot of energy density versus radial coordinate $x$.} \label{1}
\end{figure}

\begin{figure}[h!]
  \includegraphics[width=8cm]{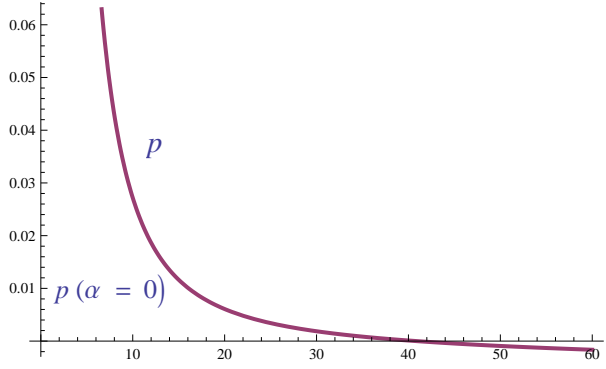}\\
  \caption{Plot of pressure versus radial coordinate $x$.} \label{2}
\end{figure}

\begin{figure}[h!]
  \includegraphics[width=8cm]{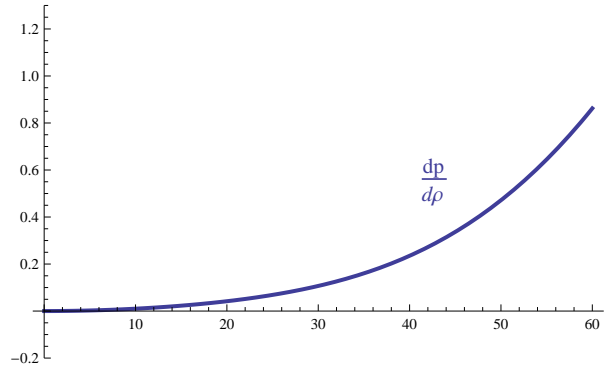}\\
  \caption{Plot of sound-speed parameter versus radial coordinate $x$.} \label{3}
\end{figure}

\begin{figure}[h!]
  \includegraphics[width=8cm]{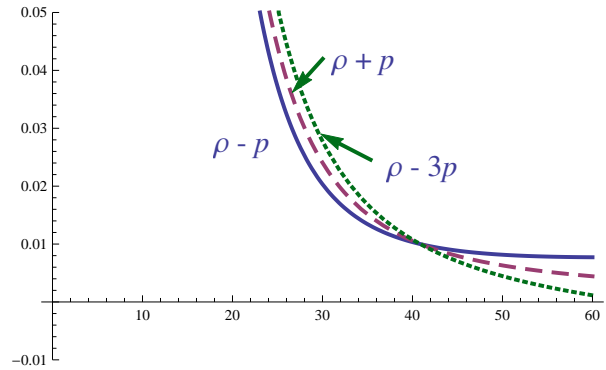}\\
  \caption{Plot of energy conditions versus radial coordinate $x$.} \label{4}
\end{figure}

\newpage
We proceed to select a set of values for the various constants in the problem in order to determine a model that harmonises best with the physical conditions. The graphs in Fig. 1--4 displayed were produced by assuming the values $ C_1 =-0.02 $, $ C_2 = -50 $, $ C_3 = -0.05 $,  $ \alpha = -150 $, $ C = 0.0002 $ and $ \beta = \alpha C $. (Observe that the use of a negative coupling constant $ \alpha $ is not novel - see for example Guo and Schwarz \cite{26}.) For these choices it is pleasing to note from Fig. 2 that a pressure-free hypersurface does indeed exist and is defined approximately by $ x = 40.8647 $ geometric units. Within this boundary we note that the pressure is positive definite, and decreases monotonically outwards from the central regions to the boundary $ r=R $. Additionally, Fig. 1  demonstrates that the density is positive and decreasing everywhere within the spherical distribution. Importantly we observe that the causality criterion $ 0 < \frac{dp}{d\rho} < 1 $ is satisfied everywhere in the interior of the star as evidenced by Fig. 3. This implies that the sound speed is never superluminal within the boundary. The energy conditions are depicted in Fig. 4. From these we infer that all the conditions: weak (solid line), strong (dashed line) and dominant (dotted line) are satisfied  within the radius of the distribution.  Using the values of the aforesaid constants as well as the boundary value $ x=40,8647 $ allows us to compute the mass as $ M= 1.51 \times 10^7 $ geometric units. Working in these same units we find that the mass to radius ratio is $ \frac{M}{R} = 369700 $ which clearly violates the Buchdahl \cite{27} limit $ \frac{M}{R} < \frac{4}{9} $ valid for stars in the Einstein general theory of relativity. This suggests that the Buchdahl upper bound may not hold in this model in EGB gravity  theory.

The dashed curve in each of  Fig. 1 and Fig. 2 reflects the situation when the EGB coupling constant is set to 0. In other words, it shows the 5-dimensional Einstein analogue. It can be seen from Fig. 2 that there is no discernible difference between the pressure profiles in the Einstein and the EGB scenarios. The density plot in Fig. 1 however, demonstrates that the radius for a positive energy density is improved by the presence of the EGB coupling constant. This implies that the gravitational field in the Gauss--Bonnet theory can sustain a greater amount of matter per unit radius as opposed to its Einstein counterpart. Note that the central singularity in this model is an artefact of the non-removable curvature singularity. A way of avoiding the singularity is to call upon a two--fluid scenario such as in a core-envelope model. This idea is popularly invoked in the standard 4-dimensional theory, see for example the deconfined quark core model surrounded by an envelope of barotopic matter of Sharma and Mukherjee \cite{28}.

Matching of the interior metric with an exterior metric such as the Boulware-Deser solution is achieved via setting
\begin{eqnarray}
\left( \frac{1}{2} C_{1} C^{2} R^{4} + C_{2} \right) ^{2} &=& 1 + \frac{R^2}{4 \alpha} \left( 1 - \sqrt{1+ \frac{ 8 M \alpha}{R^4}} \right), \label{29a} \\ \nonumber
 \\ 1 + \frac{R^2}{4\alpha} \left( 1 - \sqrt{1 + \frac{8 M \alpha}{R^4}} \right) &=& \frac{3 C_{1} C^{2} R^{4} + 8 \beta C_{1} C R^{4} + 2 C_{2} \pm C_{2} \Upsilon}{24 \beta C_{1} C R^{4}}, \label{29b} \\ \nonumber
 \\ 0 &=& 27 C_{1} ^{2} C^{4} R^{8} + 96 C_{1} ^{2} \beta C^{3} R^{6} + 8 C_{1} \left( 3 C_{2} + 32 C_{1} \beta ^{2} \right) C^{2} R^{4}  \nonumber \\
 & \quad & - 64 C_{1} C_{2} \beta C R^{2} + 4 C_{2} ^{2} \nonumber \\
 & \quad & - \left( 3 C_{1} C_{2} C^{2} R^{4} + 16 C_{1} C_{2} \beta C R^{2} - 2 C_{2} ^{2} + 2 C_{2} ^{2} \Upsilon \right) \Upsilon. \label{29c}
\end{eqnarray}
The last equation arises as a result of the vanishing of the pressure at $ r = R $. There are sufficient free parameters in the model to ensure that conditions (\ref{29a})--(\ref{29c}) are satisfied. We find that
\begin{eqnarray}
C_{1} &=& \frac{-b \pm 2 a E}{4 a} \pm \frac{1}{2} \sqrt{\frac{4 a^{3} E B \pm \left[ 4 a b e - b^{3} - 8 a^{2} f \right] }{4 a ^{3} E}}, \label{30a} \\ \nonumber
 \\ C_{2} &=& \frac{2 A - C_{1} C^{2} R^{4}}{2}, \label{30b} \\ \nonumber
 \\ C_{3} &=& \frac{ \left( A + C_{1} C^{2} R^{4} + 4 \beta C_{1} C R^{2} \left[ 1 - 3 A^{2} \right]   \right) ^{2} }{27 \beta ^{2} C _{1} ^{2} C^{2} R^{4} \left( 2 A - C_{1} C^{2} R^{4} \right) ^{2} }  + \frac{ 64 \beta C_{1} C_{2} C R^{2} - 4 C_{2} ^{2} }{432 \beta ^{2} C _{1} ^{2} C^{2} R^{4} } \nonumber \\
 & \quad & - \frac{ 9 C_{1} ^{2} C^{4} R^{8} + 96 \beta C_{1} ^{2} C^{3} R^{6} + 4 C_{1} C^{2} R^{4} \left( 16 \beta ^{2} C_{1} + 3 C_{2} \right)  }{432 \beta ^{2} C _{1} ^{2} C^{2} R^{4} }, \label{30c}
\end{eqnarray}
where
\begin{eqnarray}
A & = & \sqrt{ 1 + \frac{R^2}{4 \alpha} \left( 1 - \sqrt{1+ \frac{ 8 M \alpha}{R^4}} \right) }, \nonumber \\
a & = & 11 C^{8} R^{16} + 112 \beta C^{7} R^{14} + 240 A^{2} \beta C^{7} R^{14} - 192 \beta ^{2} C^{6} R^{12} + 768 \beta ^{2} C^{6} R^{12}, \nonumber \\
b & = & -196 A C^{6} R^{12} - 576 A \beta C^{5} R^{10} - 1152 A^{3} \beta C^{5} R^{10} - 1024 A \beta ^{2} C^{4} R^{8} - 3072 A^{3} \beta ^{2} C^{4} R^{8}, \nonumber \\
e & = & - 76 A^{2} C^{4} R^{8} - 16 C^{4} R^{8} + 1088 A^{2} C^{3} R^{6} + 1724 A^{4} C^{3} R^{6} - 128 \beta C^{3} R^{6} \nonumber \\
& \quad & + 1792 A  \beta ^{2} C^{2} R^{4} - 256 \beta ^{2} C^{2} R^{4}, \nonumber \\
f & = & 68 A C^{2} R^{4} + 640 A^{3} C R^{2} - 128 A C R^{2} - A^{5} \beta C R^{2}, \nonumber \\
h & = & 64 A^{4}, \nonumber
\end{eqnarray}
with
\begin{eqnarray}
E & = & \sqrt{\frac{2^{\frac{1}{3}} b^{2}  + \left( 3 \times 2^{\frac{1}{3}} b^{2} - 2^{\frac{10}{3}} a e + 4 a g^{\frac{1}{3}} \right) g^{\frac{1}{3}} }{3 \times 2^{\frac{7}{3}} a ^{2} g ^{\frac{1}{3}}}}, \nonumber \\
B & = &  \frac{3 \times 2^{\frac{8}{3}} a \left( e^{2} - 3 b f + 12 a h \right) + \left( 3 \times 2^{\frac{1}{3}} b^{2} g^{\frac{1}{3}} - 2^{\frac{10}{3}} a e g^{\frac{1}{3}} - 2^{\frac{5}{3}} a \left[ e^{2} - 3 b f + 12 a h \right]  \right) - 2 a g^{\frac{2}{3}} }{3 \times 2^{\frac{4}{3}} a ^{2} g ^{\frac{1}{3}}}, \nonumber
\end{eqnarray}
and
\begin{eqnarray}
g & = & 2 e^{2} - 9 b e f + 27 a f^{2} + 27 b^{2} h - 72 a e h \nonumber \\
& \quad & + \sqrt{-4 \left[ e^{2} - 3 b f + 12 a h \right] ^{3} + \left[ 2 e^{2} - 9 b e f + 27 a f^{2} + 27 b^{2} h - 72 a e h \right] ^{2} }. \nonumber
\end{eqnarray}
From (\ref{30a})--(\ref{30c}) we observe that the free parameters $ C_{1} $, $ C_{2} $, $ C_{3} $ in the model are defined in terms of $ \alpha $, $ M $, $ R $ (    and $ C $). It is interesting to note that the parameters that arise in the integration are defined in terms of physically relevant quantities: $ \alpha $ is the Gauss--Bonnet coupling constant, $ R $ is the radius and $ M $ is the mass of the star.

\section{Conclusion}

We have produced  an exact solution for a static spherically symmetric distribution of perfect fluid in the modified EGB gravity theory. The model has been studied for physical admissability and has been found to satisfy several elementary tests for physical reality. The pressure function vanishes for a particular radius and this hypersurface identifies  the boundary of the fluid. Within this boundary the pressure and energy density profiles are positive for the choice of constants made. Importantly the fluid is found to be causal as the sound speed is subluminal. The weak, strong and dominant energy conditions are found to be satisfied everywhere in the interior. Finally matching with the Boulware--Deser exterior metric is permitted. We point out that the general junction conditions for matching in EGB gravity has been considered by Davis \cite{15}. The higher order curvature terms lead to a modified set of junction conditions to be satisfied at the stellar surface. These surface equations are very different from general relativity, and their complexity makes it unlikely to easily demonstrate an exact solution. For a complete stellar model of a star in EGB gravity those boundary equations should be satisfied. In this treatment we have shown that it is possible to find bound EGB interior solutions with variable densities and pressures with desirable physical features.

\section*{ACKNOWLEDGEMENTS}
\noindent B. Chilambwe thanks the University of KwaZulu-Natal for a scholarship. B. Chilambwe and S. Hansraj thank the National Research Foundation for financial support. S. Maharaj acknowledges that this work is based upon research supported by the South African Research Chair Initiative of the Department of Science and Technology and the National Research Foundation.

\appendix*

\section{Further New Exact Solutions in the Einstein Case}

We note that equation (\ref{8c}), or (\ref{9}), admits large classes of solutions for prescribed forms of $ Z $ and $ y $. A particular 5--dimensional exact model in the Einstein case $ (\alpha = 0) $ was presented in Section IV. Other exact solutions to (\ref{9}) are possible for different forms of $ Z $. In this appendix we list a few new solutions in various categories. However, for the sake of brevity we do not study these solutions in detail.

\subsection{The form $ Z= 1 + x^n $}

With this choice for $Z$, equation (\ref{8c}) assumes the form
\begin{equation}
2x^{2-n}(x^n + 1) \ddot{y} + nx \dot{y} + (n-1) y = 0, \label{31}
\end{equation}
and has solutions which are the Legendre polynomials given by
\begin{eqnarray}
y(x) &=& C_1 \sqrt{x} {\mbox {LegendreP}} \left(\frac{-n+\sqrt{n^2 -12n + 12}}{2n}, \frac{1}{n}, \sqrt{x^n + 1} \right)  \\ \nonumber \\
& \quad & + C_2 \sqrt{x} {\mbox {LegendreQ}} \left(\frac{-n+\sqrt{n^2 - 12n + 12}}{2n} , \frac{1}{n}, \sqrt{x^n + 1} \right), \label{32}
\end{eqnarray}
where the standard Legendre functions of the first and second kind are defined, as usual, in terms of hypergeometric and gamma function. For certain values of $ n $ the Legendre functions reduce to elementary functions.
Some  of these cases are presented in Table 1.

It should be noted that the $ n = 1 $ case corresponds to the model treated by 
Dadhich et al \cite{11}, and which was shown to be equivalent to the interior Schwarzschild solution. The other  cases appear to be novel.

\begin{table}
\caption{\label{tab:table1} Forms for the potential $ y $ for specific values of the parameter $n$.}
\begin{ruledtabular}
\begin{tabular}{rc}
$n$ & Potential $y$ \\
\hline
 1 & $2C_1 \sqrt{1+x} + C_2$ \\
    2 & $ C_{1} \cos \left[ \frac{\sinh ^{-1} \left[ x \right] }{\sqrt{2}} \right] + C_{2} \sin \left[ \frac{\sinh ^{-1} \left[ x \right] }{\sqrt{2}} \right]  $ \\
    - 1 & $ \frac{ C_{1} \sqrt{1 + x} + C_{2} \left( \sqrt{x} (3 + x) - 3 \sqrt{1 + x} \sinh ^{-1} \left[ \sqrt{x} \right]  \right) }{\sqrt{x}} $ \\
    -2 & $x\left( C_1 \left(\frac{\sqrt{x^2+1}+1}{\sqrt{x^2 + 1} - 1}\right)^{-\frac{\sqrt{10}}{4}} + C_2 \left(\frac{\sqrt{x^2+1}+1}{\sqrt{x^2 + 1} - 1}\right)^{\frac{\sqrt{10}}{4}}\right) $\\
    $\frac{1}{2}$ & $ C_1 x + C_2 \left( 3x \ln \frac{\sqrt{1+\sqrt{x}}+1}{\sqrt{1+\sqrt{x}} -1} + 2(2-3\sqrt{x} )(\sqrt{1+\sqrt{x}}) \right)$ \\
\end{tabular}
\end{ruledtabular}
\end{table}

\subsection{The form $ Z= (1+x)^2 $}

The general form $ Z=(1+x)^n $ does not appear to yield elementary solutions except in the case $ n=1 $ (coinciding with the above) and $ n=2 $. In the latter case the exact solution is given by
\begin{equation}
y(x) = C_1\cos \left[ \frac{\ln (1+x)}{\sqrt{2}} \right] + C_{2} \sin \left[ \frac{\ln (1+x)}{\sqrt{2}} \right], \label{33}
\end{equation}
where $ C_{1} $ and $ C_{2} $ are integration constants.  Consequently the line element for this solution has the form
\[
ds^2 = - \left(C_1\cos \left[ \frac{\ln (1+x)}{\sqrt{2}} \right] + C_{2} \sin \left[ \frac{\ln (1+x)}{\sqrt{2}} \right]\right)^2 dt^2 + \frac{1}{(1+x)^2} dr^2 + r^2 d \Omega^2,
\]
where $ d \Omega^2 $ is the line element for the customary 3-sphere.

\subsection{The form $ Z= \frac{1}{1+x} $: 5--dimensional Finch--Skea model}

Even though the case $ Z=\frac{1}{1+x} $ is a simple form, it is worthy of special attention in its own right by virtue of the fact that this ansatz has been used by Finch and Skea \cite{20} in four dimensions to generate physically reasonable stellar models that conform to the astrophysical theory of Walecka \cite{24}, and extended by Hansraj and Maharaj \cite{21} to include charge. The Finch--Skea model was a correction of the earlier work of Duorah and Ray \cite{25}. For the 5--dimensional case the master field equation (\ref{8c}) is
\begin{equation}
2(1+x)\ddot{y} - \dot{y} + y = 0. \label{34}
\end{equation}
Note that its counterpart in 4--dimensions is given by
\begin{equation}
4(1+x)\ddot{y} -2 \dot{y} + y = 0, \label{35}
\end{equation}
where a difference in the initial coefficients is evident. The solution to (\ref{34}) is given by
\begin{equation}
y(x)=(c_2 + c_1 w) \cos w + (c_2 w - c_1) \sin w, \label{36}
\end{equation}
where $ w = \sqrt{2(1+x)} $ and $ c_1 $ and $ c_2 $ are arbitrary constants to be established by matching with the exterior Schwarzschild--Tangherlini  solution. On the other hand the Finch--Skea equation (\ref{35}) is solved by
\begin{equation}
y(x)=(c_2 + c_1 v) \cos v + (c_2 v - c_1) \sin v, \label{37}
\end{equation}
where $ v = \sqrt{1+x} $. Therefore it is clear that the physical properties of these solutions should be practically the same as there is only a difference of the factor of $ \sqrt{2} $. Clearly the change of dimension to five does not materially influence the physics of the perfect fluid.

\bibliography{basename of .bib file}

\end{document}